\documentclass[journal]{IEEEtran}

\ifCLASSINFOpdf

\else

\fi
\usepackage{soul}
\usepackage{float}
\usepackage{amssymb}
\usepackage{amsmath, amsthm, amsfonts}
\usepackage{amsfonts} % if you want blackboard bold symbols e.g. for real numbers
\usepackage{graphicx} % if you want to include jpeg or pdf pictures
\usepackage{algorithm}
\usepackage{algorithmic}
\usepackage[update,prepend]{epstopdf}
\usepackage{cite}
\usepackage[dvipsnames]{xcolor}
\usepackage{soul}
\usepackage[caption=false]{subfig}
\usepackage{etoolbox}
\usepackage{caption}
\usepackage{comment}
\usepackage{balance}
\usepackage{booktabs}
\makeatletter
\patchcmd{\@makecaption}
  {\scshape}
  {}
  {}
  {}
\makeatletter
\patchcmd{\@makecaption}
  {\\}
  {.\ }
  {}
  {}
\makeatother

\begin{document}

\title{Non-contact Vital Signs Monitoring \\ through Visible Light Sensing}
\markboth{To Appear in IEEE SENSORS JOURNAL, VOL. X, NO. X, 20XX}{Shell \MakeLowercase{et al.}: Bare Demo of IEEEtran.cls for Journals}
\author{Hisham~Abuella,~\IEEEmembership{Student~Member,~IEEE,} and Sabit~Ekin,~\IEEEmembership{Member,~IEEE}

\thanks{Copyright (c) 2019 IEEE. Personal use of this material is permitted. However, permission to use this material for any other purposes must be obtained from the IEEE by sending a request to pubs-permissions@ieee.org. (\textit{Corresponding author: Sabit Ekin}.)}
	
\thanks{A preliminary version of this work has been reported as an extended abstract in IEEE International Conference on Sensing, Communication and Networking (SECON), June 2019 \cite{secon19}.}

\thanks{H.~Abuella and S.~Ekin are with the School of Electrical and Computer Engineering, Oklahoma State University, Oklahoma, USA (e-mail:~hisham.abuella@okstate.edu,~sabit.ekin@okstate.edu).}
}

\maketitle

\begin{abstract}
This paper presents a non-contact vital signs (respiration and heartbeat) monitoring system that utilizes visible light sensing (VLS) technology. We have for the first time demonstrated the ability to wirelessly (non-contact) sense vital signs using only reflected incoherent light signals from a human subject. The VLS-based system is implemented by using simple visible light source, photodetector and data acquisition/processing unit, and is used with the developed signal processing algorithms to turn slight variations in reflected light power into accurate measurements of respiration and heart rate. To assess the accuracy of the our method, the results were compared with reliable measurements using a contact-based monitoring device (ground truth). 
More than 94\% of accuracy was observed in test results including both breathing and heartbeat rates in different scenarios as compared to the state-of-the-art baseline methods such as contact-based vitals monitoring devices. These competitive results have demonstrated that VLS-based vitals monitoring innovation is indeed a viable, powerful, attractive, low-cost and safe method. This study represents a substantive departure from the traditional ways of doing non-contact vitals monitoring methods (e.g., radio-frequency-based radars and imaging-based cameras) and is poised to make big contributions to this area. 
Since vital signs monitoring is a ubiquitous element of medicine, this work would also impact the entire health care community, from patients in their homes, to doctor's offices, to large medical institutions and industries. This technology has potential to address numerous conditions and situations in which vital signs are critical indicators such as sleep apnea and human-computer-interaction applications.

\end{abstract}


\IEEEpeerreviewmaketitle

%%%%%%%%%%%%%%%%%%%%%%%%%%%%%%%%%%%%%%%%%%%%%%
%%%%%%%%%%%% SECTION %%%%%%%%%%%%%%%%%%%%%%%%%
%%%%%%%%%%%%%%%%%%%%%%%%%%%%%%%%%%%%%%%%%%%%%%
\section{Introduction}
\IEEEPARstart{V}{ital} signs monitoring is critical for assessing the health of patients as it gives important and timely information on the patient physiological state\cite{OCAST_Ref_1,OCAST_Ref_2,OCAST_Ref_3,OCAST_Ref_4}. Conventional techniques for tracking vitals require body contact, and most of these techniques are intrusive\cite{OCAST_Ref_5,OCAST_Ref_6,OCAST_Ref_7}. Body contact sensors (e.g., electrocardiograph electrodes) can irritate or damage a patient's skin, interfere with patient treatment or comfort, provide a vector for infection and cross-contamination, and simply impede mobility. In addition, patients might feel uncomfortable (e.g., anxious, nervous, and excited) when sensors/wires are placed on their bodies. Such negative experiences can alter the respiration and heart rate of patients, producing misleading results to healthcare providers. Consequently, this has prompted the need for effective sensing methods that can wirelessly (non-contact/remote) monitor the vital signs.

Researchers have been pursuing forms of non-contact monitoring of vital signs since the 1970's\cite{OCAST_Ref_4,OCAST_Ref_8,OCAST_Ref_18,OCAST_Ref_19,OCAST_Ref_20}. Several important medical scenarios motivate a continued and determined research effort in this area; (1) cases where patients have delicate or injured skin such as low-birth weight, pre-term newborns, patients in burn units, (2) cases where wiring and ECG leads endanger or perturb the patient such as (infants with) sudden-infant-death-syndrome (SIDS) and sleep apnea, (3) general cases, which also touches financial concerns due to hospital-acquired infections (HAIs) such as cross-contamination among patients (e.g. reusable ECG leads). Non-contact vitals monitoring technology would address all of these specific problem classes.  It involves no electrodes or adhesives, no skin contact, no leads or risk of entanglement or patient discomfort, no expendables, and of course no chance of HAIs by inadequately sterilized equipment. Indeed, with sufficient development, it seems conceivable that non-contact monitoring would replace contact-based monitoring in even common medical situations, based on hygiene and convenience alone. Aside from these cases, non-contact methods are expected to more broadly impact vital signs monitoring, from in the home to the most advanced medical facilities. Because of this wide applicability, such technology has potential to address numerous conditions and situations in which vital signs are critical indicators. These include: sleep apnea, surgical procedures, SIDS, acute heart failure prediction, shock states or respiratory failure, military (defense/protection) and triage applications, disaster medicine, and smart-home health applications. Non-contact vitals monitoring has even wider applicability when it is considered outside of the umbrella of traditional medicine, for example, covert monitoring of suspects (i.e., lie detection), anxiety-monitoring for defense and homeland security, and real-time vitals monitoring of pilots, drivers (for crash avoidance) and passengers. In addition, there are numerous studies by researchers in the human-computer-interaction (HCI) community on the use of vitals data for various applications such as gaming, virtual reality and augmented reality. 

Advances in ubiquitous sensing technologies have led to intelligent systems that can monitor vital signs such as respiration (breathing) and heart rates in a non-contact modality \cite{OCAST_Ref_1,OCAST_Ref_2,OCAST_Ref_4,OCAST_Ref_8,OCAST_Ref_9,OCAST_Ref_10,OCAST_Ref_11,Vitals_VLS_Patent}. The two well-known state-of-the-art non-contact vitals monitoring methods are based on radio-frequency (RF) (radar)\cite{OCAST_Ref_8,OCAST_Ref_4,OCAST_Ref_20,OCAST_Ref_18,OCAST_Ref_19,H_Old_Ref4,H_Old_Ref6,H_Old_Ref8} and imaging (camera)\cite{OCAST_Ref_9,Ref_5_2,OCAST_Ref_37,OCAST_Ref_38,OCAST_Ref_39,OCAST_Ref_41,H_Old_Ref1,H_Old_Ref2}. 
 An initial proof of concept using electromagnetic (EM) RF signals (radar) was successfully carried out by the pioneers in this field \cite{OCAST_Ref_8,OCAST_Ref_18}. In\cite{OCAST_Ref_8}, Chen et al. developed a sensitive microwave life-detection which can be used to find human subjects buried earthquake rubble or hidden behind various barriers.
 On the other hand, advances in image processing allowed researchers to develop a imaging-based photoplethysmography (PPG) vital signs monitoring system\cite{OCAST_Ref_37,OCAST_Ref_38,OCAST_Ref_39,OCAST_Ref_40,OCAST_Ref_41}, which uses digital camcorders/cameras with ambient light as the illumination source to measure the measuring changes in light absorption\cite{OCAST_Ref_34}. In\cite{Ref_5_2}, a method for the extraction of respiration from imaging-based measurements taken from the top-view of an incubator for critically-ill or premature infants was developed. In addition to RF- and imaging-based methods, infrared and vibration sensors were used to estimate the vitals of a  human subject \cite{Geophone_Sensor_Vitals,PIR_Sensor_Vitals}. In \cite{Geophone_Sensor_Vitals}, Jia et al. used geophones to sense the vibrations on the bed caused by human heart and respiration. In \cite{PIR_Sensor_Vitals}, Erden et al. utilized multiple infrared sensors to estimate respiration rate for sleep apnea detection.
 
 These technologies offer optimistic performance, but their practicality is still uncertain since they face continued technical and implementation challenges. Visual-based sensing introduces tremendous privacy concerns since it relies on closed-circuit or worse, networked cameras \cite{OCAST_Ref_13}. On the other hand, RF-based methods may lead to electromagnetic interference (EMI) in critical electrical circuits (e.g., life-saving medical equipment) \cite{OCAST_Ref_14,OCAST_Ref_15}. The latter can also present safety concerns in terms of long-term exposure and short distance \cite{OCAST_Ref_16,OCAST_Ref_17}, and therefore are not universally accepted, particularly in hospitals (See Appendix for further discussion). In addition, the vitals monitoring methods that use infrared and vibration sensors have only been used to estimate vital signs in a specific scenario such as subject laying on a bed. Further, multiple sensors are needed in infrared based method \cite{PIR_Sensor_Vitals}, and vibration sensors need to be in-contact with the bed's frame. At the end, all these current wireless (non-contact) sensing methods are at their infancy and require significant enhancements to improve their capabilities and be adopted in commercial medical product.
 
In this study, we present a novel vital signs monitoring system\footnote{Patent pending: S. Ekin and, H. Abuella, "System  and  method  of  non-contact  vital sign  monitoring through light sensing", US Patent No. 62/639,524.}, a non-contact sensing technology that monitors breathing and heart rates through visible light sensing (VLS). 
Visible light is still in some ways an untapped portion of the spectrum, which has sparked a revolution \cite{NSF_Ref_38,pathak2015visible}. It offers immense potential for sensing and communication areas because of its ubiquitous, safe and low-cost features. It is present everywhere and is gaining significant interest as a medium for wireless communication and only recently for sensing applications\cite{pathak2015visible,NSF_Ref_40,NSF_Ref_41,NSF_Ref_42}. 
These innovations are being realized secondary to the advances in solid-state lighting, sensor technologies and signal processing. This breakthrough creates a new range of exciting as well as critical applications \cite{NSF_Ref_38}. For example, 1) Philips is transforming lighting infrastructures into a method for providing localization services \cite{NSF_Ref_60}, 2) companies, like PureLiFi, LVX, and OLEDCOMM, provide Internet connectivity through LEDs at data rates comparable to WiFi \cite{NSF_Ref_61,NSF_Ref_62}, and 3) Disney has developed a new generation of interactive toys using visible light \cite{NSF_Ref_63}. In addition, in recent years, visible light communication, an evolving communication technology that utilizes visible light signals for wireless data transmission, has received great attention\cite{pathak2015visible,NSF_Ref_64,NSF_Ref_65}

The main contributions of this study are as follows:
\begin{itemize}
\item  A novel VLS-based non-contact vital signs monitoring system has been presented (Fig.~\ref{fig:System_Model}).

\item  The proposed vitals monitoring system has been implemented (an opto-electronic system design) including hardware and software components.  

\item The signal processing algorithms have been developed to turn raw data into actionable vitals information. 

\item  The developed vitals monitoring system as a whole has been evaluated and iteratively optimized with human subjects in a realistic setting under different lighting conditions and body positions.  

\item  Performance of the system has been measured against standards produced by globally accepted contact-based off-the-shelf medical vitals monitoring equipment. 

\end{itemize}

 The rest of the paper is organized as follows. Motivation and significance of the proposed non-contact vitals monitoring method is provided in Section~\ref{sec:motivation}. Next, system design, theory of operation, signal processing algorithms used to estimate breathing and heart rates, and implementation details are given in Section~\ref{sec:sys_model}. Following that, the experimental evaluation and results are presented in Section~\ref{sec:Evaluation}. Finally, conclusions and future research directions are discussed in the last section.

%%%%%%%%%%%%%%%%%%%%%%%%%%%%%%%%%%%%%%%%%%%%%%
%%%%%%%%%%%% SECTION %%%%%%%%%%%%%%%%%%%%%%%%%
%%%%%%%%%%%%%%%%%%%%%%%%%%%%%%%%%%%%%%%%%%%%%%
\section{Motivation and Significance of the VLS-based Non-contact Vitals Monitoring Method}
\label{sec:motivation}
Historically, it was believed that the small variations in the received signal (in both RF- and imaging-based non-contact methods) were because of noise and unpredictable reflections and wave scattering from the environment. However, as the technology in sensing and signal processing improved, researchers discovered that some of these variations were sourced from humans' physiological movements (heartbeats and respiration), and that one could extract this information. With this at the forefront of our thinking, and by recognizing the decades of high-sensitivity photodetector development from the fiber-optic industry, we hypothesized that it was possible to extract the physiological data from visible light signal variations and to monitor the vital signs remotely. 

We develop a VLS-based vitals monitoring system, a non-contact sensing technology that monitors breathing and heart rates. The basic principle of our method is that a visible light signal (generally ambient light) is transmitted toward the human body (subject), where it is amplitude-modulated by the periodic physiological movements and reflected back to the receiver. The VLS receiver captures the reflected signal and processes it to extract the vital sign components, which are the tiny periodic variations in the received signal. In contrast, the RF-based method uses the RF signal and physical movement of the body \cite{OCAST_Ref_20,2018_wifi_Vitals}, while visual-based method uses the ambient light (visible light) and the skin absorption level of the light\cite{OCAST_Ref_37,OCAST_Ref_38}. To summarize, the VLS receiver (photodetector) captures the reflected signal while processing algorithms extract the vital sign signals as illustrated in Fig.~\ref{fig:System_Model}.  

Our developed system is more advantageous compared to the RF methods in terms of EMI and safety and does not introduce any privacy concerns as in imaging-based systems. Our developed system is safe, secure, power-efficient and can be used to monitor the vitals of patients in any healthcare environment.

%%%%%%%%%%%%%%%%%%%%%%%%%%%%%%%%%%%%%%%%%%%%%%
%%%%%%%%%%%% SECTION %%%%%%%%%%%%%%%%%%%%%%%%%
%%%%%%%%%%%%%%%%%%%%%%%%%%%%%%%%%%%%%%%%%%%%%%

\section{System Design and Implementation}\label{sec:sys_model}

\begin{figure}[t]
\includegraphics[width=0.45\textwidth]{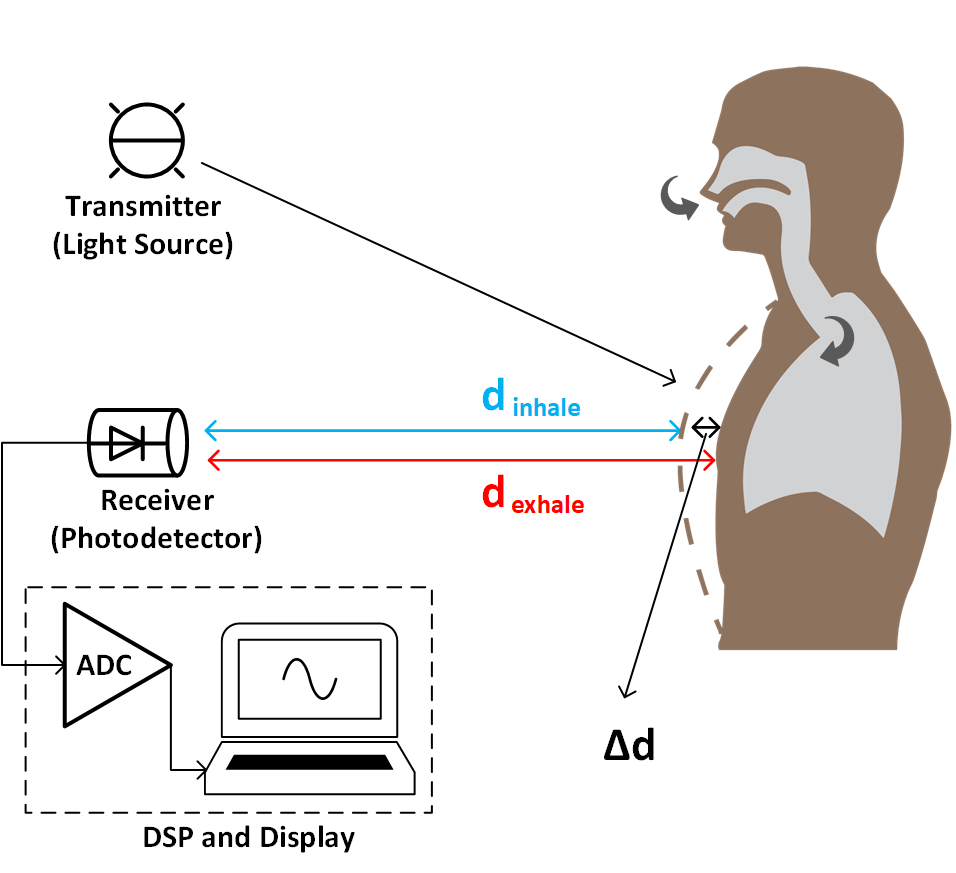} 
\centering
\caption{System design: an illustration of the VLS-based non-contact vital signs monitoring system.}
\label{fig:System_Model}
\end{figure}

In this section, our proposed VLS-based non-contact vital signs monitoring system is presented. First, we discuss the general system design and its main components. Second, we provide the theory of operation of our VLS-based system and explain the light signal variation and the system parameters might affect a VLS based system. Third, we introduce our signal processing algorithms used to estimate the breathing and heart rates from the received light signal. Finally, we provide details for the system implementation including both hardware and software components.  

\begin{figure}[t]
\includegraphics[width=0.48\textwidth]{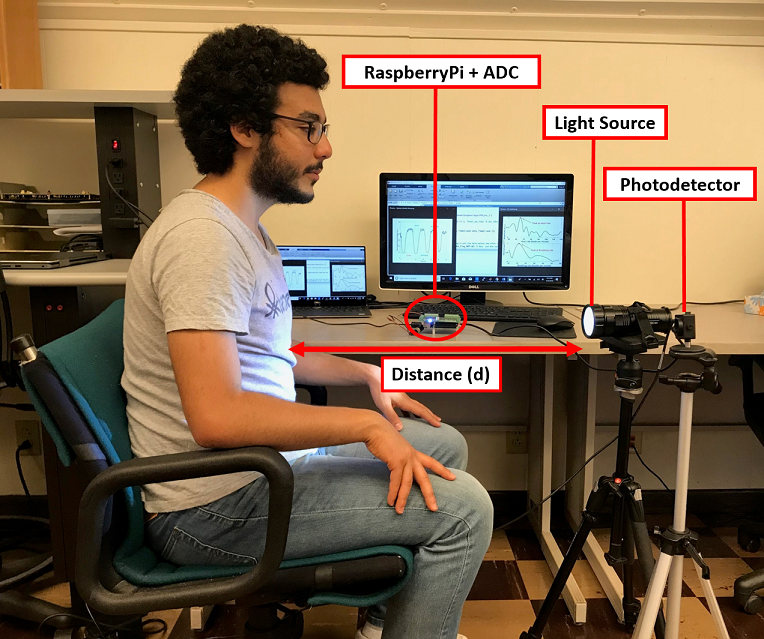} 
\centering
\caption{The experimental setup of  VLS-based non-contact vital signs monitoring system (sitting scenario).}
\label{fig:Hardware_Setup_sitting}
\end{figure}

Our envisioned VLS-based non-contact vital signs monitoring system is depicted in Fig.~\ref{fig:System_Model}, where the $\Delta d$ is the distance difference between inhale and exhale positions.
The system mainly consists of 1) photodetector as receiver, 2) light source as transmitter, and 3) digital signal processing (DSP) unit and display, which includes the analog to digital conversion (ADC) unit. 
Our experimental set up is given in Fig.~\ref{fig:Hardware_Setup_sitting}. Although, this setup was developed for proof-of-concept purposes with low cost units (see Section \ref{sec:Implementation}), our basic DSP algorithm was able to extract vital signs information with very high accuracy (see Section \ref{sec:Evaluation}).

\begin{figure*}[ht]
\includegraphics[width=0.9\textwidth]{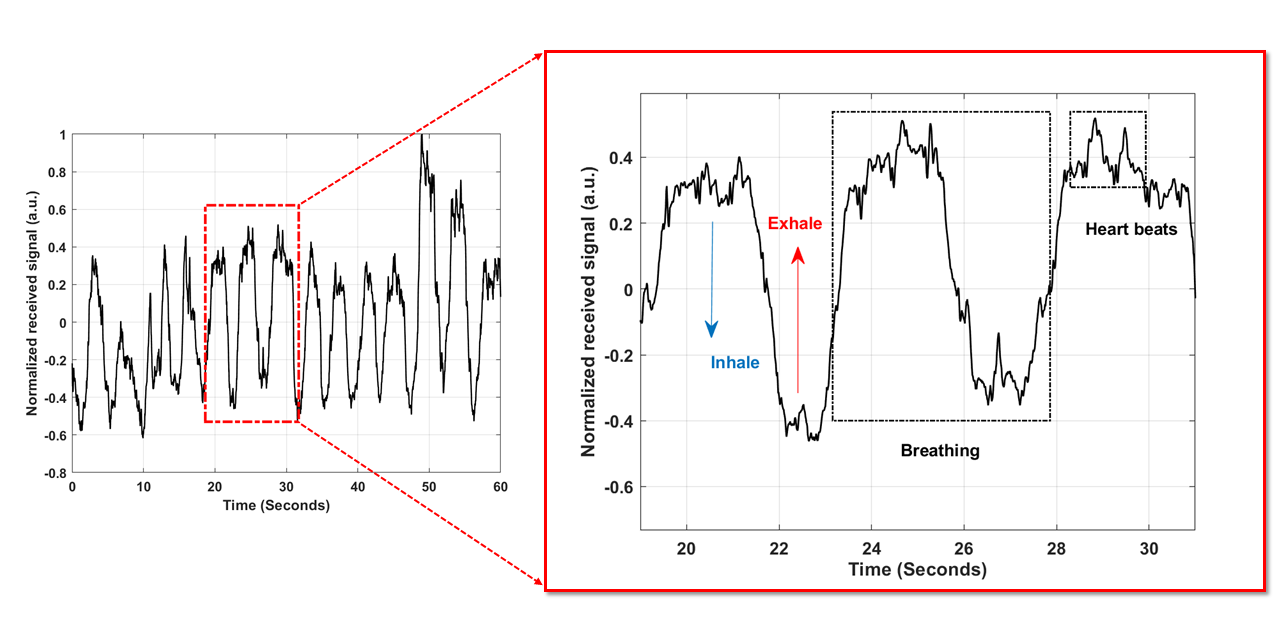}
\centering
\caption{Raw data (normalized): received light signal power.}
\label{fig:raw_data}
\end{figure*}

To illustrate how the received light signal power varies with the vital signs, let us consider the example from our experimental setup (Fig.~\ref{fig:Hardware_Setup_sitting}), where the test subject sits facing the VLS system. 
When the subject inhales, his chest expands and gets closer to the device; and when he exhales, his chest contracts and gets farther away from the device. As mentioned, because the signal power and the distance to a reflector are proportionally related, the VLS system can track a person's breathing and heartbeats. Fig.~\ref{fig:raw_data} shows the signal power of the captured reflection as a function of time.  
In addition, as shown in Fig.~\ref{fig:raw_data}, the high power  corresponds to the inhale position (crest) while the lower power  corresponds to the exhale position (trough) as the light signal travels a longer distance, hence the low power is received. Note that the heartbeat signal is modulated on top of the breathing signal.

\subsection{Theory of Operation}
\label{sec:Theory}

To better understand the proposed VLS-based vitals monitoring system, it is very helpful to know a little about the theory of operation of the VLS.
The visible light channel has been studied intensively in the context of wireless communications \cite{VLC_State_of_Art,pathak2015visible}, one of the well-known and adopted channel models is the Lambertian model\cite{VLC_First_channel_Model}. 
This model captures the effect of different variables and parameters at the receiver such as the transmitter power, distance between light source and photodetector, optical size of photodetector, field of view (FoV), and irradiance and incidence angles. The received signal power of visible light for the Lambertian light source is given by 
\begin{equation}
   \label{eq:Lambertian_Channel1}
                  P_r(t)=\frac{(n+1)A_R P_t}{2\pi [d(t)]^{\gamma}}\cos^n(\phi)\cos(\theta) , \forall\theta< \phi_{1/2},
\end{equation}
where $P_t$ is the power of transmitted signal, $\gamma$ is the empirical path-loss exponent, $A_R$ is the optical detector area, and $\phi$ and $\theta$ are irradiance and incidence angles, respectively.
In addition, $d(t)$ is the distance between the transmitter and the receiver, $\phi_{1/2}$ is the semi-angle at half-power of the light source, and $n$ is the order of the Lambertian model and is given by
\begin{equation}
n=-\frac{\ln(2)}{\ln(\cos\phi_{1/2})}.
\end{equation}
\begin{comment}
In our case the light source and the receiver are static and at same height, therefore we can have
     \begin{equation}
     \label{eq:Theta_PHI}
                   \theta = \phi,
  \end{equation}
 where  $ 0 <\theta < \phi_{1/2} $. Using~\eqref{eq:Theta_PHI},~\eqref{eq:Lambertian_Channel1} can be further simplified as
\begin{equation}
 \label{eq:Lambertian_Channel2}
%\nonumber
 P_r(t)=\frac{(n+1)A_R}{2\pi [d(t)]^{\gamma}}\cos^{n+1}(\theta).
 \end{equation}
Finally, in order to derive $P_r(t)$ in terms of $d(t)$, we further simplify \eqref{eq:Lambertian_Channel2} by defining a constant $K$ as:  
\begin{equation}
%\nonumber
K = \frac{(n+1)A_R \cos^{n+1}(\theta)}{2\pi}.
\end{equation}
\end{comment}  
This well-known model underpins visible light-based sensing and communication research \cite{pathak2015visible,NSF_Ref_67}.
Furthermore, the relation between received signal power and distance can be readily inferred from the Lambertian channel model as
\begin{equation}
   \label{eq:Lambertian_Channel3}
                  P_r(t) \propto[d(t)]^{-\gamma},
  \end{equation}
where $P_r(t)$ is the received signal power as a function of time $t$, $d(t)$ is the distance between the transmitter and the receiver. The path-loss exponent $\gamma$ (typ. 1-5) depends on the environment conditions. 

In our setup, the relation between received power and distance is used as: $ P_r(t) = K [d(t)]^{-\gamma}$, where $K$ is a constant value that incorporates all losses in the system, equal to $-111.2$ dB and $\gamma$ is equal to 3.238 (which are shown in Fig. \ref{fig:Distance_Effect} and Fig. \ref{fig:Position_Effect}. Further, note that the value of $\Delta d$ is highly dependent on subject's size (i.e., volume), age, gender and even health condition, hence can vary from person to person. In our experiments, the observed $\Delta d$  was varying  between 0.5-2 cm for breathing, and is expected to be much less for heart rate (e.g., $<$0.2 cm). Nonetheless, it is worth to mention that the most important step in our sensing method is to extract periodicity information from the received signal, which would represent the vital signs. The values of $K$, $\gamma$ and  $\Delta d$ would be highly dependent on environment, utilized hardware and subject's size, respectively. Therefore, the values provided above are for representative purposes, and should not taken as ground truth.

Notice that the received power changes with the distance, as expected. The longer the distance between the transmitter (light source, e.g., LED) and receiver (photodetector), the lower the received power is.  Most importantly, variations in the distance $d(t)$ with certain periodicity result in similar variations in the received power with the same periodicity.  
In other words, the above equation shows that one can identify variations in $d(t)$ due to breathing (inhaling and exhaling) and heartbeats, by measuring the resulting variations in the amplitude (power) of the received (reflected) signal. These small movements are used to extract the rates for breathing and heartbeats.
This relation between the received power and distance serves as the foundation of our invention. Experimental results prove the viability of the general approach.

\begin{figure}[t]
\includegraphics[width=0.45\textwidth]{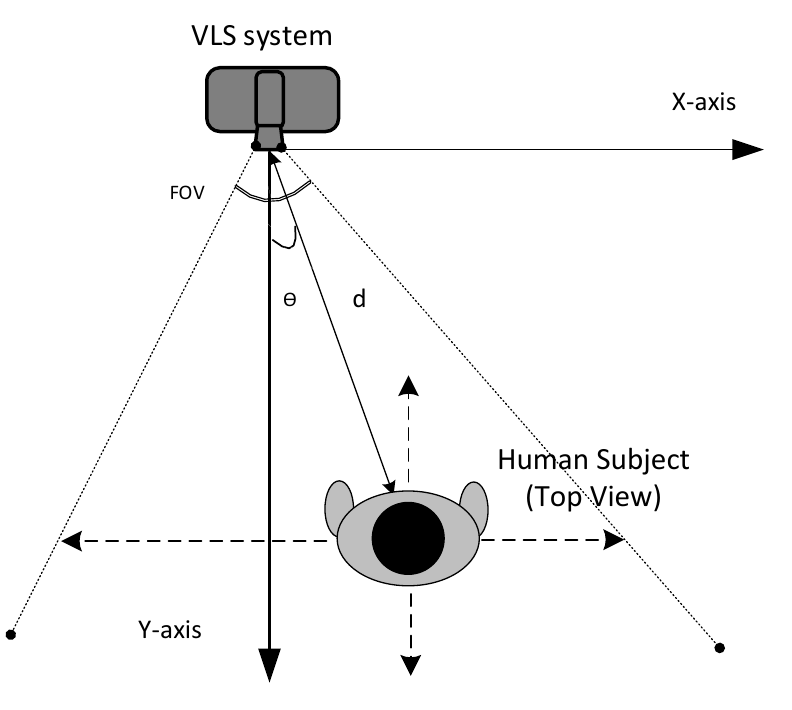}
\centering
\setlength{\belowcaptionskip}{-8pt} 
\caption{Top view of the system to show the effect of distance and incident angle on the reflected received signal power.}
\label{fig:Distance_angle_System}
\end{figure}

As depicted in Fig.~\ref{fig:Distance_angle_System}, the position of the subject can affect the received signal power $P_r$, since the distance $d$ and the incidence angle\footnote{It is known from the Lambertian channel model that the incidence angle $\theta$ can impact the received power with the relation of $\cos(\theta)$ factor, where the maximum power is received at $\theta=0^o$.} between the subject and the photodetector change. Note that the impact of the distance on the system estimation accuracy is provided in Experimental Evaluation and Results (Section~\ref{sec:Evaluation}).

\subsection{Signal Processing Algorithms}
\label{sec:Estimation_Algorithms}

Once of the most critical parts in such non-contact sensing is to extract the frequency tones of breathing and heart rates from a noisy time domain light signal using signal processing algorithms to monitor human vitals. Note that similar approaches (algorithms) have been used in  RF-based vital signs monitoring studies\cite{H_Old_Ref6,H_Old_Ref8,OCAST_Ref_20}.

\begin{figure}[t]
\includegraphics[width=0.26\textwidth]{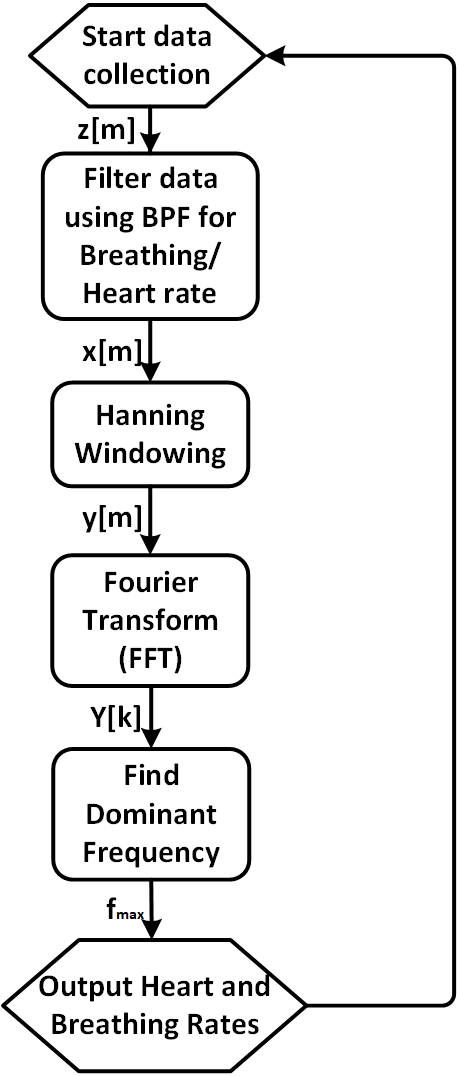} 
\centering
\caption{Flow diagram of the vitals (heart and breathing rates) estimation algorithm.}

\label{fig:Flow_Diagram}
\end{figure}

\begin{figure}[t]
\includegraphics[width=0.49\textwidth]{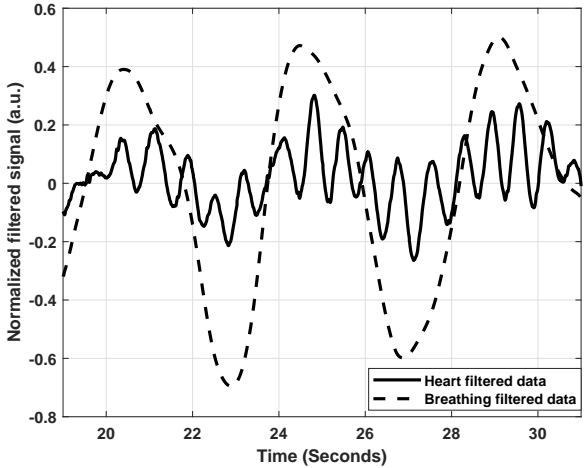}
\centering
\caption{The output of the filters used for breathing and heart rates.}

\label{fig:Filtered_Data}
\end{figure}

It is known that breathing and heartbeats are periodic motions, hence we can extract the frequencies (rates) of breathing and heartbeats by performing a Fourier transform (an FFT). In addition, having different periodicities allows separate filtering operations for breathing and heartbeat.
The flow diagram of our algorithm is shown in Fig.~\ref{fig:Flow_Diagram}, where the same data is processed for breathing and heart rates seperately. 
We estimate the breathing (or heart) rate by detecting the highest frequency tone in the band of normal breathing (or heart) rate for an adult human.  This can be tailored to other age ranges.  First, we filter the raw data received (See Fig.~\ref{fig:raw_data}) from the photodetector with an infinite impulse response (IIR) Chebyshev Type-2 band pass filter (BPF) of appropriate passband range for adult breathing (or heart) rates. We filter the frequency domain signal around (10-60) breaths per minute and (30-200) beats per minute for breathing and heart rates, respectively. The IIR filtering is done as:
\begin{equation}
   \label{eq:Filteringing}
                 x[m]=\sum_{i=0}^{P}b_i z[m-i] -\sum_{j=1}^{Q}a_j x[m-j],
  \end{equation}
where $z[m]$ is the digital time domain window (dataset) data, $x[m]$ is the digital time domain window (dataset) of the filtered data, and $b_i$ and $a_i$ are the IIR feedforward and feedback filter coefficients\footnote{Filter coefficients would vary depending on desired filter type and bandpass frequencies. In this work, we used two IIR filters. First filter used for heart rate estimation has coefficients of $ b =[0.0010   -0.0077    0.0262   -0.0517    0.0642   -0.0517    0.0262   -0.0077    0.0010]$ and $ a= [1.0000   -7.8359   26.8895  -52.7799   64.8128  -50.9871   25.0939   -7.0643    0.8709]$. 
Second filter used for  breathing rate estimation has coefficients of $b=[0.0007 0   -0.0013 0    0.0007]$ and $a=[1.0000   -3.9247    5.7781   -3.7820    0.9286]$}. $P$ and $Q$ represent the orders of the IIR feedforward and feedback filters, respectively.   
The effect of filtering the data is shown in Fig.~\ref{fig:Filtered_Data}.

After filtering the data, we multiply each dataset with a Hanning window to remove the spectral leakage before the FFT process as follows:
\begin{equation}
   \label{eq:Windowing}
                 y[m] = W_{hanning}[m]x[m],
  \end{equation}
where $y[m]$ is the input digital signal to the FFT process, $W_{hanning}[m]=\sin^2(\frac{\pi m}{N})$ and $N$ is the size of the dataset window. Since the FFT works well with infinite data, if we want to use it accurately with finite data we have to remove the leakage introduced by bypassing the convolution of the frequency domain data with the rectangular window infinite frequency response.

\begin{figure}[t]
\includegraphics[width=0.49\textwidth]{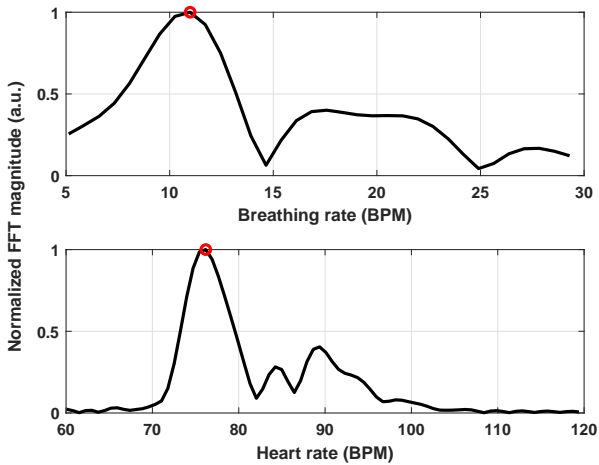}
\centering
\caption{The frequency domain of the data after filtering for breathing and heart rates (BPM: beats  per  minutes for heart  rate and breathing  per  minutes for  breathing rate).}
\label{fig:Freq_Domain_Data}
\end{figure}

After this step, we take the FFT (which is a faster implementation of the discrete frequency transform (DFT)) of the time domain data as follows:
\begin{equation}
   \label{eq:FFT}
                 Y[k]=\sum_{m=0}^{N-1} y[m]e^{\frac{-j2\pi km}{N}},
  \end{equation}
where $Y[k]$ is the output of the FFT process which represents the different frequency tones ($f_k$s) available in the signal. 
Finally, we detect the highest power tone ($f_{max}$) in the region of interest as follows (Fig.~\ref{fig:Freq_Domain_Data}):  
\begin{equation}
   \label{eq:Freq_Max}
        k_{\max} = \underset{k}{\arg \max}~|Y[k]|.
  \end{equation}
Finally, the frequency where the maximum FFT value observed can be obtained by $f_{\max} = f[k_{\max}]$. 
We carry out these steps on each dataset and at the end we calculate the average of the breathing (or heart) rate during the whole test.

\subsection{Implementation}\label{sec:Implementation}

In this section, we present the hardware used to implement the system prototype. Then, we discuss the software and the measurements parameters used. Finally, we present the devices and ground truth which is used as a reference for our system performance evaluation.

\subsubsection{Hardware}
\label{sec:Hardware}
We have prepared our experimental setup by using a Thorlabs PDA100A photodetector \cite{PDA_100A}, a RaspberryPi miniature computer \cite{RaspeberryPi},  an ADC (PiPlate) circuit \cite{PiPlate}, an off-the-shelf fixed power LED light source, and display unit as shown in Fig.~\ref{fig:Hardware_Setup_sitting}.  In the presented setup, the human subject will be sitting in front of the light source and the photodetector, where the reflected signal is received at the photodetector and sent to the PiPlate for digital conversion. Finally, the digital data is stored at the RaspberryPi for further analysis and processing. 

\subsubsection{Software} 
We implemented our data collection software by using Python scripting language run in PiPlate and RaspberryPi. DSP algorithms described in Fig.~\ref{fig:Flow_Diagram} have been implemented in both Matlab and Python. The duration for each measurement is 60 seconds and then the data is separated into windows for estimating the vital rates. The distance between human subject and VLS system is 40 cm unless stated otherwise. The algorithm FFT size (window size) is 2048 (around 20 seconds for ADC sampling rate ($F_s$) = 100 samples/second).

 \subsubsection{Ground Truth}
 \label{sec:Ground_Truth}
The contact-based (off-the-shelf) devices used to compare our VLS-based system performance are Spire Stone \cite{Spire_OffShelf} and Kardia for ECG measurements \cite{Kardia}. For each measurement taken using our proposed system, a reference value for the heart and breathing rates is taken using these FDA approved devices.

%%%%%%%%%%%%%%%%%%%%%%%%%%%%%%%%%%%%%%%%%%%%%%
%%%%%%%%%%%% SECTION %%%%%%%%%%%%%%%%%%%%%%%%%
%%%%%%%%%%%%%%%%%%%%%%%%%%%%%%%%%%%%%%%%%%%%%%

\section{Experimental Evaluations}
\label{sec:Evaluation}
In this section, the experimental environment, evaluation and comparison results are presented to provide the performance of VLS-based vitals monitoring system. First, we discuss the systems parameters such as distance and incidence angles and how they are affecting the performance of the estimation. Then, the performance of the vitals monitoring system under different scenarios are shown. Finally, the comparison results are presented.

In the experimental setup and environment; 1) the tests have been conducted with multiple participants, 2) the light source power is constant (no flickering within the frequency range of interest (0.08 - 3 Hz),  3) subjects avoid major limb motions\footnote{It is know that major limb motions (subject moves and/or walks) can severely impact the accuracy of most of the vital signs monitors including contact-based devices (e.g., pulse oximeters for monitoring heart rate, chest bands for monitoring breathing) cannot provide accurate estimates when the user walks or moves \cite{OCAST_Ref_20, lanata2010comparative, petterson2007effect}. Data from such motion intervals can easily be discarded as limp motion is aperiodic \cite{OCAST_Ref_20}.} during the measurement, 4) the monitoring system starts giving reliable estimations after 30 seconds of measurements\footnote{Similar settling time is needed for off-the-shelf vitals monitoring devices as well.}, 5) all subjects were asked to wear a grey color shirt for fair comparison of the results\footnote{Impact of clothing material and its color will be considered as future work as discussed in Section \ref{sec:conclusion}.}, and 6) measurements results are per minute, i.e., beats per minutes (BPM) for heart rate and breathing per minutes (BPM) for breathing rate. Last but not least, in addition to sitting position, two additional experimental setups (scenarios) were evaluated: 1) subject stands in front of the VLS system, such scenario can be  applicable in the airport security check points, and 2) subject lays down (sleeping) and VLS system is above the subject and directed to the chest, such scenario can be applicable in the hospital setting.

We calculate the mean ($\hat{X}_{VLS}$) and variance ($\sigma^2_{VLS}$). In addition, we have the reference estimate ($X_{ref}$) using the FDA-approved devices as discussed in Ground Truth subsection (in Section \ref{sec:sys_model}). We measure the performance of our system by using two different parameters: 1) percentage of absolute error in the vitals estimation, calculated as: $100 \times \frac{\mid \hat{X}_{VLS}-X_{ref} \mid }{\hat{X}_{ref}} $, and 2) the variance of the vitals estimation ($\sigma^2_{VLS}$) during the 60 seconds measurements.

\subsection{System Parameters Analysis}

\begin{figure}[t]
\includegraphics[width=0.49\textwidth]{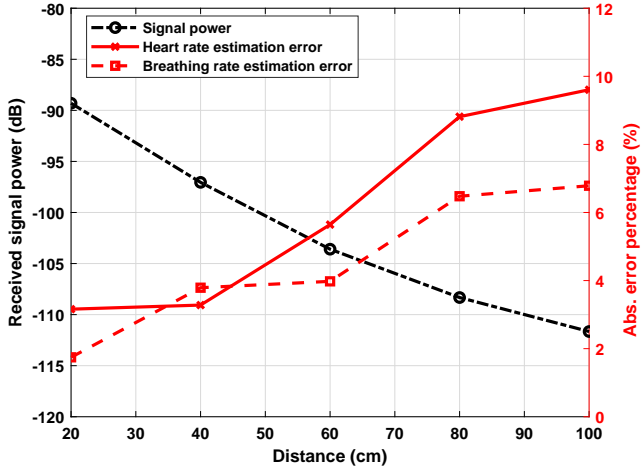}   
\centering
\caption{Received signal power, and heart \& breathing rate error percentages for different distances based on the  average of ten measurements for one subject.}
\label{fig:Distance_Effect}
\end{figure}

In Fig.~\ref{fig:Distance_Effect}, we investigate the effect of the distance between the VLS system and the human subject. As expected, as the distance increases the received signal decreases. Such decrease in the received power can impact the performance of the system, particularly in detection of tiny fluctuations due to heart beats. As one can observe from Fig.~\ref{fig:Distance_Effect}, although the estimation accuracy of heart and breathing rates are comparable at the short distance, the accuracy of heath rate decreases faster than the accuracy of breathing rate as the distance increases. Such observation is attributed to the truth that heart beats are much smaller movements than breathing (see Fig. \ref{fig:raw_data}).

\begin{figure}[t]
\includegraphics[width=0.49\textwidth]{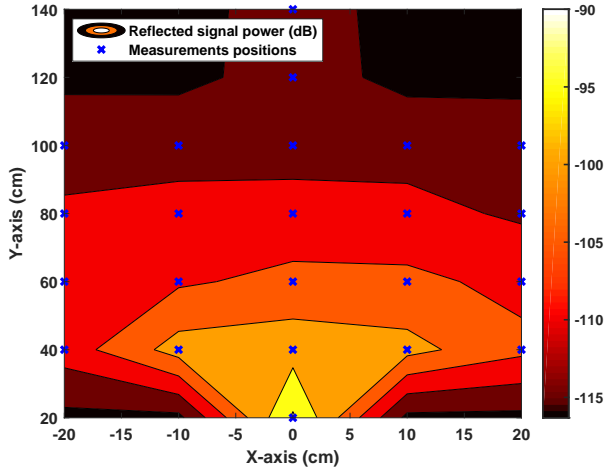}   
\centering
\caption{The received signal power at different positions from the photodetector and the light source by averaging 6 seconds measurements.}
\label{fig:Position_Effect}
\end{figure}

The photodetector's FoV is critical in the performance measurement. 
In Fig.~\ref{fig:Position_Effect}, the heatmap of the received signal power which is reflected from the human target is shown, where VLS system is positioned at the origin (0,0) cm. As the distance and incidence angles increase, the reflected received signal decreases, so does the received signal power, which decreases the estimation accuracy as shown in Fig.~\ref{fig:Distance_Effect}.

\begin{figure}[t]
\includegraphics[width=0.44\textwidth]{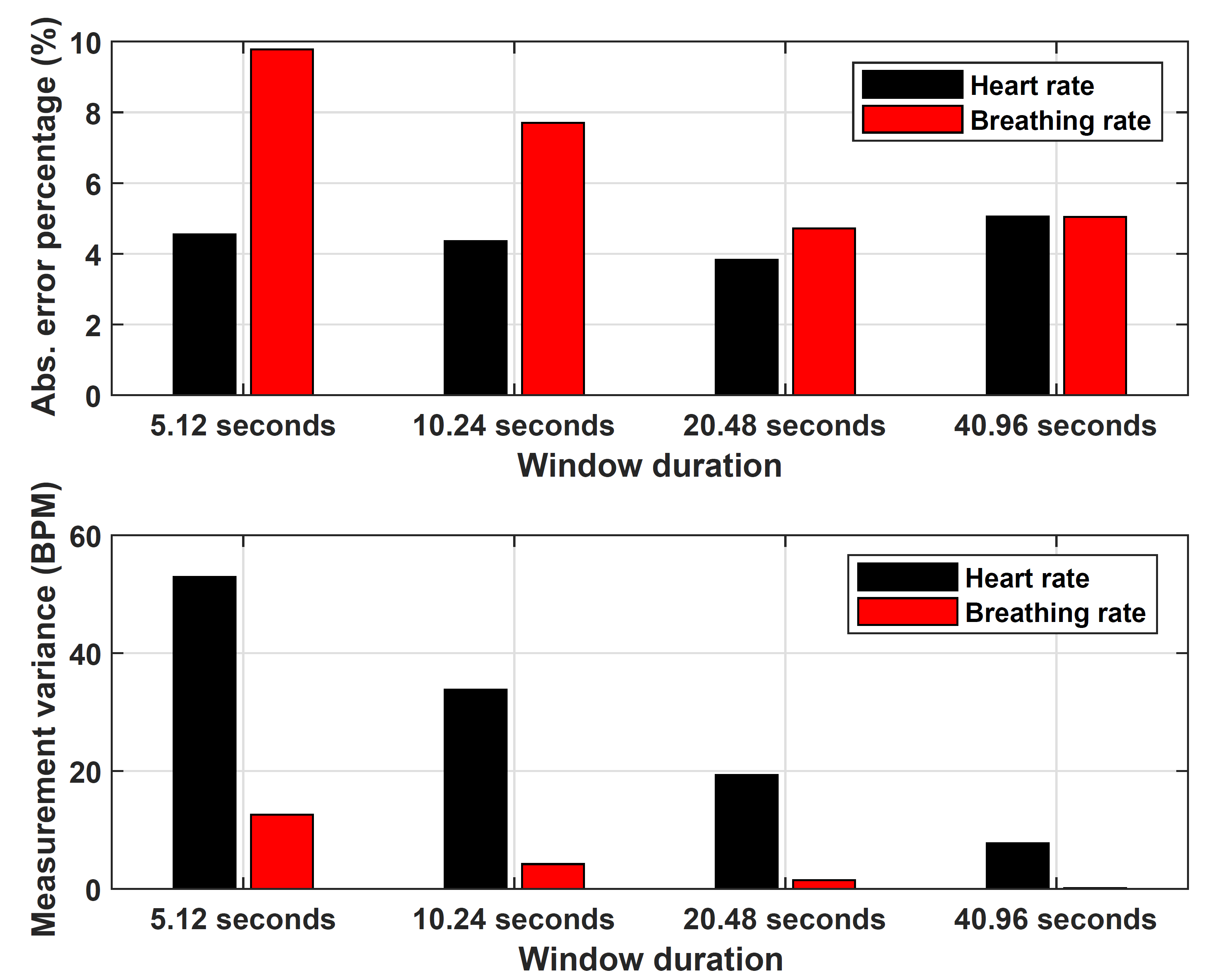}
\centering
\caption{Heart and breathing rates error percentages and estimation variances for different data window (sample) size. VLS-based vital monitoring is compared with Spire Stone and Kardia based on the average of ten measurements for one subject.}
\label{fig:FFT_Effect}
\end{figure}

\begin{figure}[t]
\includegraphics[width=0.49\textwidth]{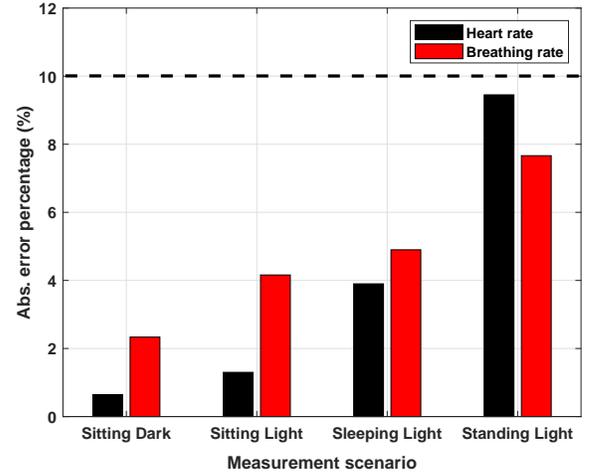}
\centering
\caption{Heart and breathing rates error percentages in different scenarios for a single subject (FFT size=2048). VLS-based vital monitoring is compared with Spire Stone and Kardia based on the average of ten measurements for one subject.}
\label{fig:Different_scenarios_Hisham}
\end{figure}

Another important parameter that is worth to investigate its impact on the performance is the sample size used for vitals estimation algorithm. In Fig.~\ref{fig:FFT_Effect}, we present the effect of different the number of sample sizes used for the estimation algorithm in terms of absolute error percentage and measurement variance. The estimation error decreases as the data window (sample) size increases. This improvement is attributed to the fact that as the window size is increased the frequency resolution between the different frequency bins at the output of FFT decreases. 
Although a larger sample size is not needed in heart beats, breathing rate would require larger sample size due to having a lower rate (BPM).  
In addition, as the window size increases the estimation variance decreases as expected.
We observe the best performance in both heart and breathing rates estimation results for a data window size of approximately 20 seconds (20.48 seconds)
Also, the distance of 40 cm is chosen in all the evaluation tests.
It is evident that if the widow size is too large, it can lead to an increase in the interference and noise from the other body movements, which can increase the estimation error.

\begin{figure*}[t]
\includegraphics[width=0.8\textwidth]{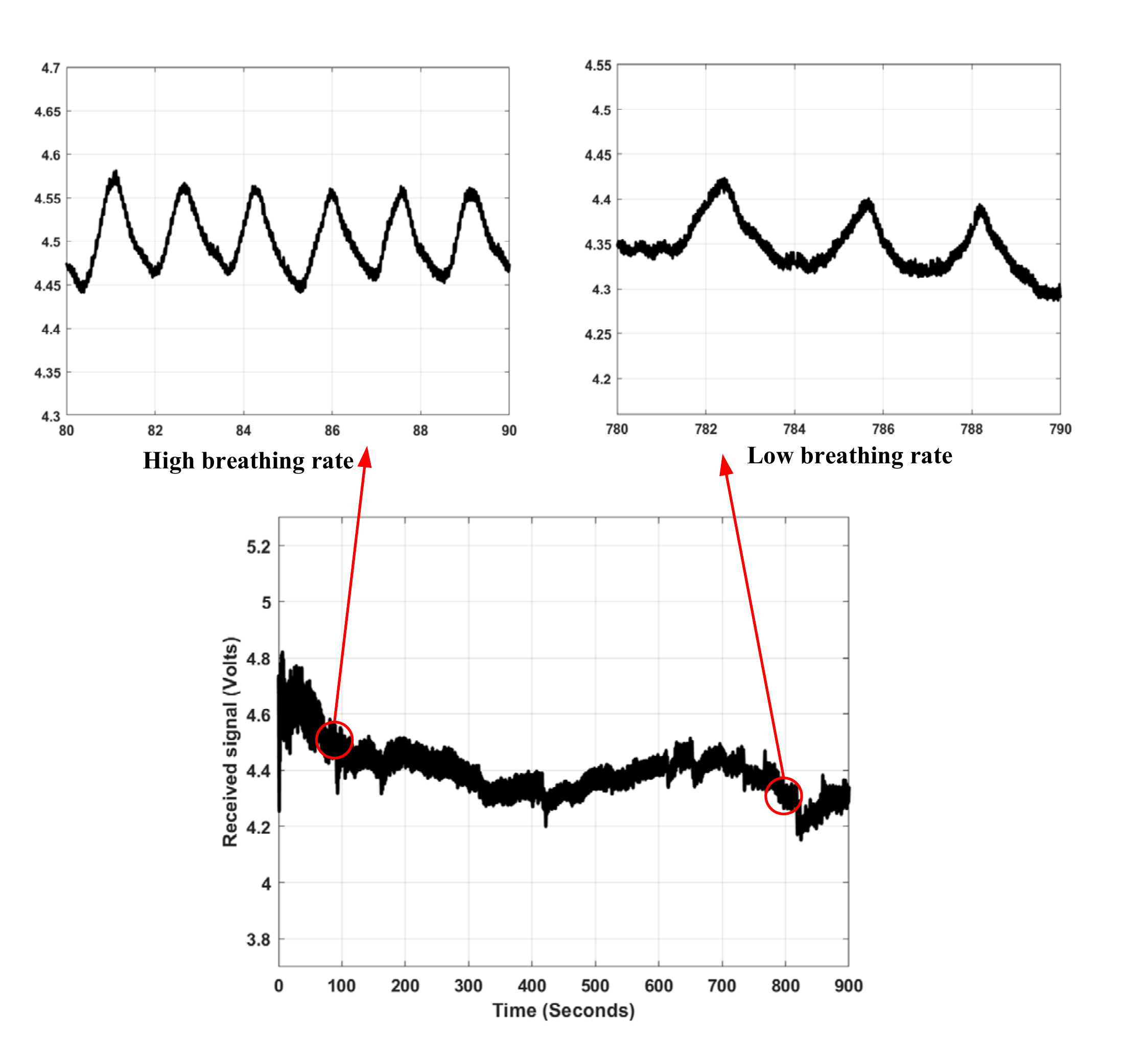}  
\centering
\caption{Received signal level for 15 minutes after a physical exercise.}
\label{fig:After_exercise}
\end{figure*}

\subsection{Performance in Different Scenarios}

In Fig.~\ref{fig:Different_scenarios_Hisham}, we present performance of our VLS-based vitals monitoring system in different scenarios to show how the system can be used in different positions without dramatically changing its accuracy. We consider three positions for the subject (sitting on a chair, sleeping on a table, standing in front of the system). In addition, two lighting conditions are considered to study the effect of the lighting condition of the room. As expected, when measurements are taken in normal lighting conditions, the system accuracy is less for the same system parameters in the low lighting conditions (dark\footnote{Measurements in dark environment were conducted to show the limits of the current system setup and algorithm as this environment would be the ideal case where there is no interference from other light sources.}). This can be explained by the fact that the noise and interference is lower in the low ambient lighting. 

\begin{figure}[t]
\includegraphics[width=0.49\textwidth]{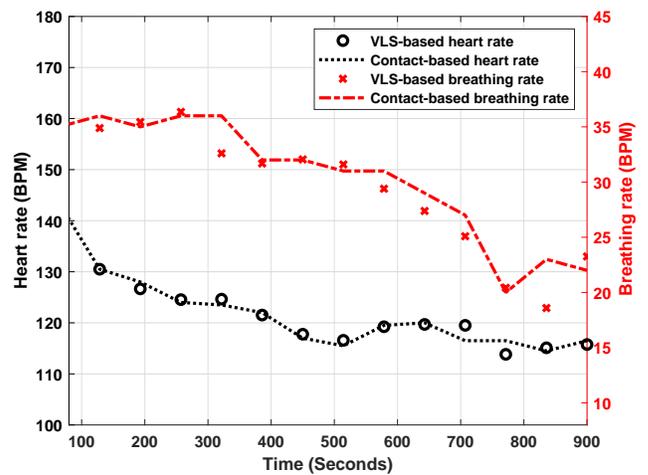}  
\centering
\caption{Heart and breathing rates estimation results for 15 minutes after a physical exercise. VLS-based vital monitoring is compared with Spire Stone and Kardia.}
\label{fig:Error_After_exercise}
\end{figure}

Intuitively, it is evident that the worst performance would be observed when the human target is standing in front of the sensing system due to unwanted subject movements during the standing scenario (Fig. 12). Overall, we observe less than 10\% error in all scenarios. In order to minimize the effect of the movement on the performance, additional software (e.g., adaptive signal processing tools) and hardware modifications (e.g., more sensitive sensors, optical lenses and filters) could be done to have a robust sensing system to dynamic environment and body movements. These improvements will depend on the scenario and the setup as discussed in the RF-based vitals monitoring systems in \cite{RELAX_RF_Vitals,Random_Body_Movement_Cancellation}.

In another scenario, the human subject was asked to perform a physical exercise and then the heart and breathing rates were measured using our VLS-based system for 15 minutes after the subject stopped exercising to monitor how vital rates and received signal decrease gradually. As shown in Fig.~\ref{fig:After_exercise}, a decrease in the breathing rate during the measurement can be noticed visually without any signal processing. A comparison of measured breathing and heart rates between the results obtained from our VLS-based system and the off-the-shelf vitals monitoring devices are presented in Fig.~\ref{fig:Error_After_exercise}, where results are matching for most of the measurement time. Nevertheless, it is worth to note that the operating range of BPFs for heart and breathing rates where updated to match the estimated increase in both of heart and breathing rates for this scenario.

\subsection{Comparisons (Multiple subjects)}
\begin{figure}[t]
\includegraphics[width=0.49\textwidth]{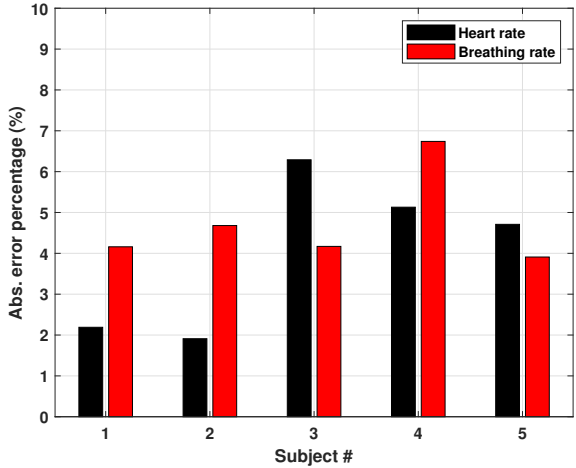}
\centering
\caption{Heart and breathing rates error percentages for multiple subjects in case of sitting in normal lighting scenario (each subject had 10 measurements with 60 seconds each). VLS-based vital monitoring is compared with Spire Stone and Kardia based on the average of ten measurements for one subject.}
\label{fig:Different_Subjects}
\end{figure}

In order to further  evaluate the system performance, multiple subjects\footnote{The statistics of the subjects were as follows: 4 male and 1 female, between (60-100) kg in weight and (1.5-1.8) m in height, and age between (25-35).} have been used and the results are compared with contact-based devices (ground truth). As shown in Fig.~\ref{fig:Different_Subjects},  the system performance was evaluated using five different subjects and around 50 measurements. The worst accuracy observed for measurements including both heart and breathing rates was 94\%. It is worth mentioning that the off-the-shelf devices \cite{Spire_OffShelf} are assumed to be 100\% accurate.

%%%%%%%%%%%%%%%%%%%%%%%%%%%%%%%%%%%%%%%%%%%%%%
%%%%%%%%%%%% SECTION %%%%%%%%%%%%%%%%%%%%%%%%%
%%%%%%%%%%%%%%%%%%%%%%%%%%%%%%%%%%%%%%%%%%%%%% 
\section{Conclusions and Future Research Directions}\label{sec:conclusion}
%\subsection{Conclusions}
In this paper, we presented the idea of using VLS in the non-contact vitals monitoring. By using off-the-shelf light source and photodetector, we collect the reflected visible light signal from a human subject. Our developed VLS-based system was able estimate heart and breathing rates accurately by applying signal processing algorithms (filtering, FFT and windowing). In order to evaluate the system performance, we compared the performance results with the off-the-shelf contact-based devices. Our developed system is able to extract vitals rate information with very high accuracy.
The developed technology is still in its infancy and many improvements in hardware and signal processing algorithms should be achieved before it could be used clinically used and tested. It will be a useful heart-monitoring method that can overcome the difficulties of conventional heart monitors requiring physical contact with the patient which will make it a valuable non-contact-based vitals monitoring system in various applications in medical facilities, security checks (detecting any change in the human vitals to detect the subject mood). This study is the \textit{first attempt} at creating a VLS-based non-contact vital signs monitoring method and investigating its potential. It serves as \textit{proof-of-concept} to validate our original hypothesis of using VLS for vitals monitoring. Finally, we presented our analysis on the limitations of different technologies that are used in non-contact vitals monitoring and how VLS can fit and add to these technologies.

% \subsection*{\textcolor{red}{Future Research Directions}}
Since VLS-based vitals monitoring is a newly developed method, much investigation (as future directions) is required to further quantify its limitations and to test its performance in different real-world scenarios, and generally improve its utility as much as possible. There can be other scientific questions that need answers. The most important question is whether this sensing modality can be feasible in practical and dynamic environments. That question broadly frames numerous specific questions that demand research for answers. Some examples are: 1) What are the quantitative opto-electronic relations between light intensity, receiver sensitivity, and the properties of the subject under test (e.g. clothing, skin color, blankets and bedding)? 2) What is the required dynamic range and signal-to-noise ratio of the opto-electronic VLS hardware system design to produce a reliable measurement? 3) What other algorithms are suitable for processing raw reflected light measurements and turning them into reliable measurements? 4) What is the quantitative effect of other light sources that may interfere, and how may they be mitigated? Finally, improving the software and hardware components and using an advanced VLS setup would attract a lot of interest. This work will inspire new thinking about how VLS-based vitals monitoring can impact many important applications.

%%%%%%%%%%%%%%%%%%%%%%%%%%%%%%%%%%%%%%%%%%%%%%
%%%%%%%%%%%% SECTION %%%%%%%%%%%%%%%%%%%%%%%%%
%%%%%%%%%%%%%%%%%%%%%%%%%%%%%%%%%%%%%%%%%%%%%% 
\section*{Acknowledgment}
This work was done with the help of current and previous students of Wireless Communication and Sensing Research Lab (WCSRL) in the School of Electrical and Computer Engineering at Oklahoma State University: Caleb G. Teague, Habeeb Idrees, and Muhammed E.Oztemel. Special thanks to Dr. John O'Hara, and Amit Kachroo  for their valuable comments and suggestions to improve this paper. Moreover, we thank all the volunteers who helped in testing the system performance.

%%%%%%%%%%%%%%%%%%%%%%%%%%%%%%%%%%%%%%%%%%%%%%
%%%%%%%%%%%% SECTION %%%%%%%%%%%%%%%%%%%%%%%%%
%%%%%%%%%%%%%%%%%%%%%%%%%%%%%%%%%%%%%%%%%%%%%% 
\appendices

\section{Limitations of RF- and Imaging-based Vitals Monitoring Methods} 
 \label{sec:remarks}
In the following, we discuss some of our underlying advantages over using VLS over RF- and imaging-based technologies (the most well-known non-contact methods) in vitals monitoring and their potential implications on practical implementation. 
\subsection{RF-based Vitals Monitoring}
One of the key challenges that impacts reliability in using wireless RF signals is extraction of vital signs due to the fact that the RF environment is very dynamic and any motion in the environment affects the signal which complicates data extraction \cite{OCAST_Ref_4}. 
RF signals can also easily penetrate through a human body, and transmit power level in an RF signal is a safety concern \cite{OCAST_Ref_16}. To minimize the impact of RF signals exposure on humans, the transmit power level in wireless devices is strictly controlled and regulated by government agencies, e.g., Federal Communications Commission (FCC) regulations on Wireless Medical Telemetry Systems \cite{OCAST_Ref_31,OCAST_Ref_32,OCAST_Ref_33}. Although it is claimed by researchers that the transmit power on RF-based monitoring systems will be safe, we have found no formal study investigating their impact on health in case of continuous and long-term exposure by multiple devices, which remains a concern. 

Further, with regard to RF sensing, EMI is a continuing concern. For over 20 years now, numerous studies have quantified the effects of EMI in special cases where certain devices impact certain medical instruments (e.g. cell phone's effect on infusion pumps)\cite{OCAST_Ref_14}.  In that time, apparently no conclusion has been reached that RF-producing devices are universally safe or harmful in medical facilities.  Nevertheless, numerous studies have demonstrated measurable interference between medical and other devices, particularly cell phones\cite{OCAST_Ref_15,NSF_Ref_23}.
Last but not least, many networking devices (e.g., WiFi routers, smart phones, Internet of Things (IoT) devices) work at similar frequencies, and the list is growing, making it more and more difficult to allocate spectrum for RF health monitors. Although researchers have been working on this technology for a few decades, there is no device available on the market due to the challenges and concerns mentioned above.

\subsection{Imaging-based Vitals Monitoring} 

It is known that imaging-based vitals monitoring is susceptible to motion-induced signal corruption, and overcoming motion artifacts is one of the most challenging problems. Mostly, the noise level falls within the same frequency band as the physiological signal of interest, thus making linear filtering with fixed cut-off frequencies ineffective\cite{OCAST_Ref_37}. In addition, camera-based vital signs monitoring is challenging for people with darker skin color\cite{OCAST_Ref_36,OCAST_Ref_40}. Image processing algorithms require high computational power, which makes camera-based methods power-inefficient and inappropriate for power-limited systems. Finally, and probably most importantly, camera-based systems introduce great privacy and security issues. Cameras are regularly mounted in “public” areas of hospitals for general security.  However, their use in operating rooms, patient rooms, and bathrooms is strongly discouraged by security and privacy experts\cite{NSF_Ref_21}. The simplest and most effective solution is to avoid the collection of patient images altogether by employing technologies that operate without cameras.

%%%%%%%%%%%%%%%%%%%%%%%%%%%%%%%%%%%%%%%%%%%%%%
%%%%%%%%%%%% SECTION %%%%%%%%%%%%%%%%%%%%%%%%%
%%%%%%%%%%%%%%%%%%%%%%%%%%%%%%%%%%%%%%%%%%%%%% 
%Where the bibliography will be printed

\bibliographystyle{IEEEtran}
\bibliography{Biblo}

\begin{IEEEbiography}[{ \includegraphics[width=1in,height=1.25in,clip]{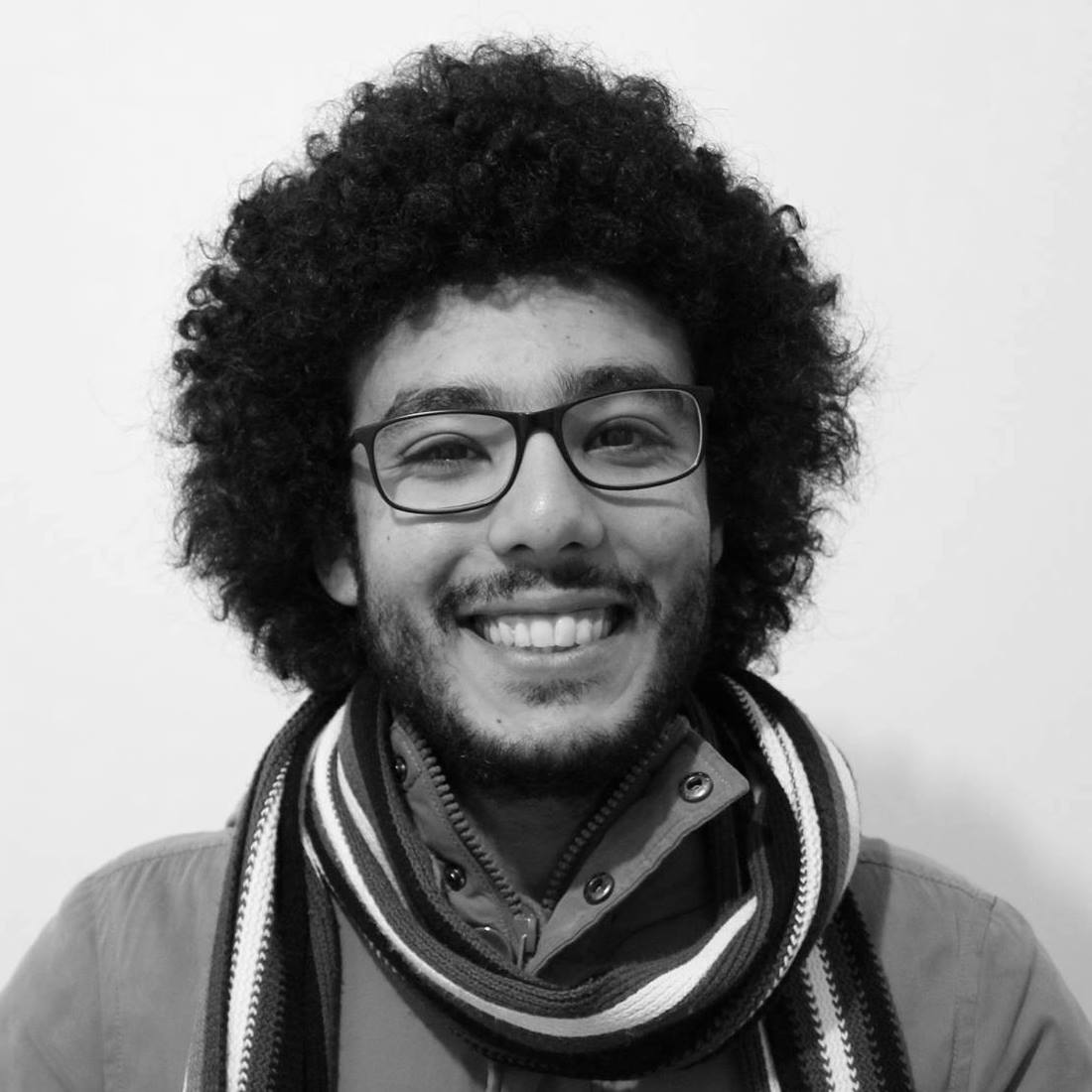}} ]{Hisham Abuella} received the B.Sc. degree in Communications and Electronics Engineering from Ain Shams University, Cairo, Egypt, in 2013. He worked as a Digital System Design Engineer in Varkon Semiconductor (Wasiela) Company, Cairo, Egypt. In Fall 2014, he joined Istanbul Sehir University as a Research Assistant for his M.Sc. degree in Electronics and Computer Engineering from  Istanbul Sehir University, Turkey. Lastly, he joined Oklahoma State University as a Graduate Research Assistant to pursue his Ph.D. study at the School of Electrical and Computer Engineering in Spring 2017. He is currently working with Dr. Sabit Ekin at Wireless Communications and Sensing Research Lab (WCSRL). He was an engineering intern in Qualcomm Inc., San Diego, CA during the summer of 2019. His current research interests include light sensing and communication, wireless communication systems design using SDRs, visible light Sensing applications, and machine learning and DSP algorithms for wireless communication systems.
\end{IEEEbiography}

\begin{IEEEbiography}[{\includegraphics[width=1in,height=1.25in,clip]{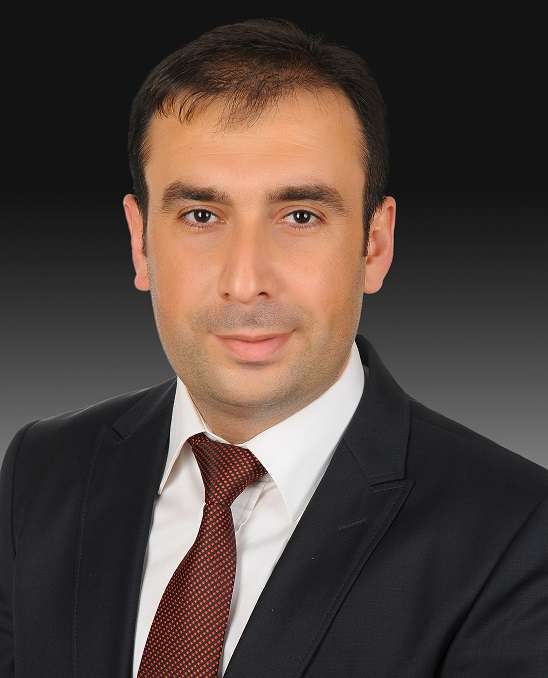}}]{Sabit Ekin (M'12)}  received the B.Sc. degree in electrical and electronics engineering from Eski\c sehir Osmangazi University, Turkey, in 2006, the M.Sc. degree in electrical engineering from New Mexico Tech, Socorro, NM, USA, in 2008, and the Ph.D. degree in electrical and computer engineering from Texas A\&M University, College Station, TX, USA, in 2012. He was a Visiting Research Assistant with the Electrical and Computer Engineering Program, Texas A\&M University at Qatar from 2008 to 2009. In summer 2012, he was with the Femtocell Interference Management Team in the Corporate Research and Development, New Jersey Research Center, Qualcomm Inc. He joined the School of Electrical and Computer Engineering, Oklahoma State University, Stillwater, OK, USA, as an Assistant Professor, in 2016. He has four years of industrial experience from Qualcomm Inc., as a Senior Modem Systems Engineer with the Department of Qualcomm Mobile Computing. At Qualcomm Inc., he has received numerous Qualstar awards for his achievements/contributions on cellular modem receiver design. His research interests include the design and performance analysis of wireless communications systems in both theoretical and practical point of views, interference modeling, management and optimization in 5G, mmWave, HetNets, cognitive radio systems and applications, satellite communications, visible light sensing, communications and applications, RF channel modeling, non-contact health monitoring, and Internet of Things applications.
\end{IEEEbiography}

% that's all folks

\end{document}